\PassOptionsToPackage{bookmarks={false}}{hyperref}
\documentclass[conference,times,mathptm,psfig,]{IEEEtran}
\usepackage{blindtext, graphicx}
\usepackage{epstopdf}
\usepackage[utf8]{inputenc}
\usepackage{listings}
\usepackage{amssymb}
\usepackage{amsmath}
\usepackage{mathtools}
\usepackage{algorithm} 
\usepackage{algpseudocode} 
\usepackage{color}
\usepackage{caption}
\usepackage{subcaption}
\usepackage{enumitem}
\usepackage{nohyperref}
\usepackage{geometry}
\begin{document}

\title{Resilient UAV Formation for Coverage and Connectivity of Spatially Dispersed Users}

\author{Yuhui Wang and Junaid~Farooq\\
Department of Electrical and Computer Engineering, \\
University of Michigan-Dearborn, Dearborn, MI 48128 USA. \\
{Emails: \{ywangdq, mjfarooq\}@umich.edu}
}

\markboth{}{}

\maketitle

\maketitle

\begin{abstract}
Unmanned aerial vehicles (UAVs) are a convenient choice for carrying mobile base stations to rapidly setup communication services for ground users. Unlike terrestrial networks, UAVs do not have fiber optic back-haul connectivity except when they are tethered to the ground, which restricts their mobility. In the absence of back-haul, e.g., in remote areas, emergency situations, or in battlefields, there is a need to ensure connectivity among UAVs in addition to coverage of ground users for creating local area networks. This paper provides a distributed and dynamic approach for UAV formation-based control for coverage and connectivity of spatially dispersed users. We use flocking dynamics as a guide to constructing tailored formations of UAVs on the fly. Simulation results demonstrate that if sufficient aerial base stations are available, the proposed approach results in a strongly connected network of UAVs that is able to provide both a backhaul and fronthaul network. The approach can be further extended to create multi-tier extra-terrestrial networks to cater for large-scale applications.
\end{abstract}


\IEEEpeerreviewmaketitle

\section{Introduction}
Unmanned aerial vehicles (UAVs) are now becoming widespread in a range of smart city applications such as package delivery, policing, transportation, etc. UAVs are also considered to play a key role in the next-generation of wireless networks (i.e., 6G and beyond) \cite{uav_survey}\cite{5G_with_UAVs}, where it can support and enhance existing cellular infrastructure to connect the unconnected. UAVs carrying base stations (BSs) can be crucial in providing communication services in certain situations such as in disaster-struck areas and battlefields \cite{iobt_twc}. In fact, tethered drones are already providing connectivity is emergency situations, e.g., AT\&T cell on wings (COWs) were used in recent hurricane Ida in Louisiana to restore LTE coverage to cellular users~\cite{hurricane_Ida}. However, these operations are on a limited scale and typically supplement existing cellular networks by relying on already available back-haul links~\cite{5g_backhaul}. 

\begin{figure}[t]
    \centering
    \vspace{-0.0in}
    \includegraphics[width=\linewidth]{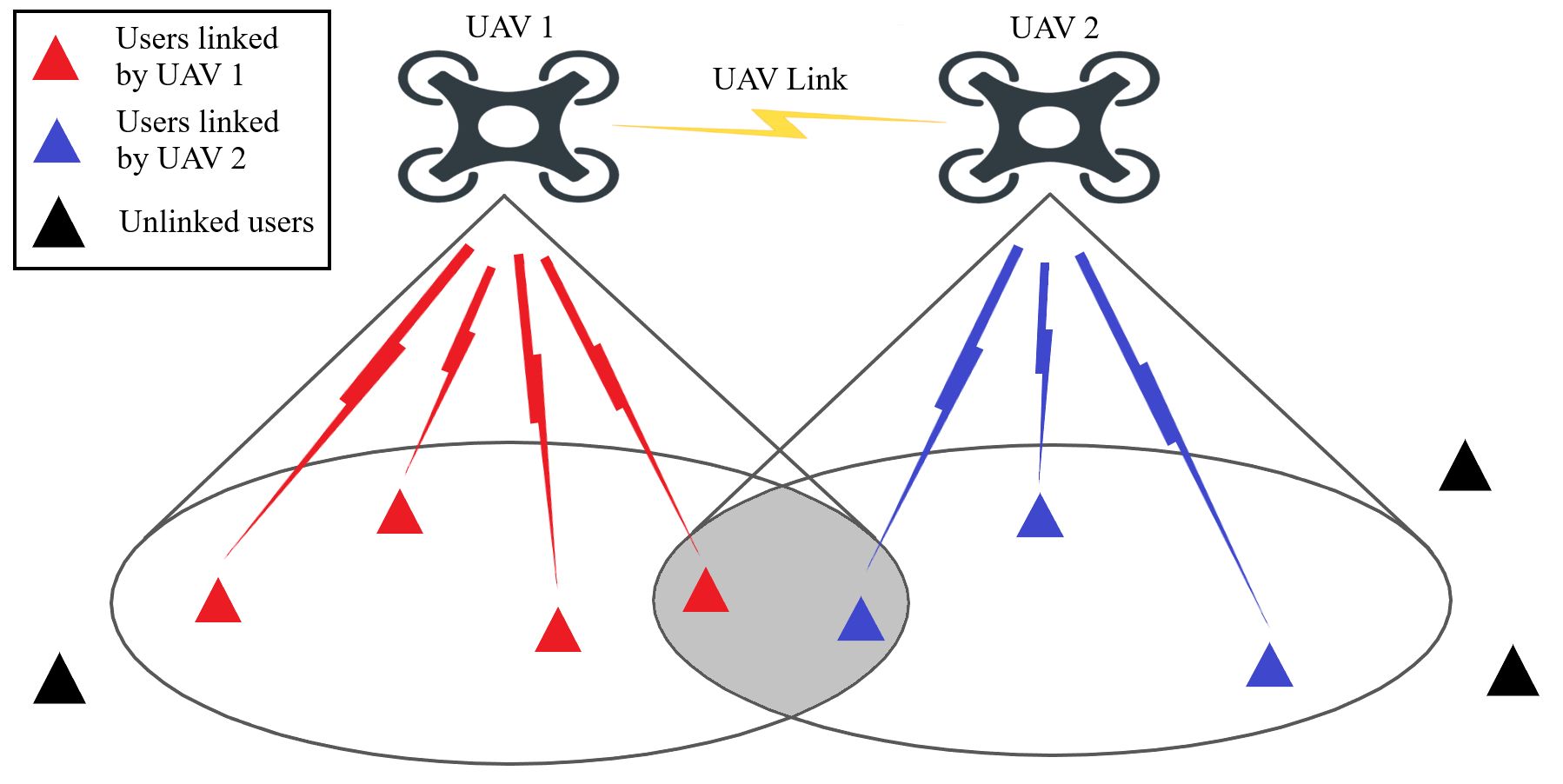} \vspace{-0.1in}
    \caption{Illustration of coverage and connectivity for two UAVs. The UAVs remain in close proximity of each other and also connect to ground users within their coverage range. \vspace{-0.1in}}
    \label{fig:J}
\end{figure}

To scale up the deployment of aerial platforms for creating local area networks, there is a need to develop adaptive strategies that are able to maintain UAV connectivity while ensuring the ground users are reachable. Fig.~\ref{fig:J} illustrates a simple scenario where two UAVs are connected to ground users within their communication range while being in close proximity to each other. This enables the formation of a local network in the absence of a back-haul. However, an extended relay network of UAVs can also connect to the wider back-haul networks via tethered links, satellite communication, or other terrestrial networks. The key challenge is to design and dynamically achieve a UAV network formation that is tailored to the locations of ground users.

The problem of creating a formation that covers users and also ensures close proximity of the UAVs is an NP-hard problem, particularly in cases when the underlying users are located in geographically dispersed clusters. In many ways, it is similar to the facility location problem in Operations Research and supply chain management~\cite{facility_location}, where facility centers need to be positioned at locations that best serves the demands of consumers. However, the additional constraint here is to ensure that facilities are close enough to create a lattice structure, which makes the problem intractable.

We use a completely distributed and dynamic approach to tackle this problem inspired from swarming or flocking dynamics in robotics and control literature. 
UAVs achieve the desired formations by operating in multiple modes including goal navigation, network connectivity establishment and user coverage optimization, which provide a natural and holistic approach to solve such multi-layer problems. 


\section{Related Work} \label{Sec:Related_work}

Existing works in the literature on UAV placement use static optimization algorithms for selecting optimal UAV locations based on various goals and objectives, e.g., maximal coverage~\cite{3d_placement}, minimizing coverage holes~\cite{3d_placement_hole}, \cite{coverage_control}.
Most works assume the availability of back-haul networks or use ground communication infrastructure as a supplement to UAV networks. Traditional operations research based approaches such as disk covering and bin packing have also been investigated for UAV drones~\cite{optimal_coverage_mozaffari}. However, they are unable to ensure connectivity among them. We have considered a completely standalone UAV system for both coverage and connectivity purposes.

A preliminary version of the framework has been presented in~\cite{connectivity_resilience}\cite{resilient_connectivity_tccn} where a flocking based control algorithm creates a formation that can provide both coverage to the users while keeping the UAVs connected~\cite{multi_layer}. However, it is only limited to scenarios where the ground users are in close proximity to each other. In other words, it does not support more complex user distributions especially where the users are spatially dispersed. In this work, we have enhanced the framework with a multi-modal system that can ensure that UAVs can participate in coverage of clusters as well as in connecting different clusters in a distributed fashion. Furthermore, the proposed dynamic approach is naturally able to adapt to changes that occur in the network over time as opposed to static optimization approaches that are not resilient to failures and attacks.

\section{System Model} \label{Sec:Sys_Model}
Consider a set of ground users referred to as mobile smart devicess (MSDs) $\mathcal{M} = \{1, \ldots, M\}$, that are placed arbitrarily in $\mathbb{R}^2$ and a set of UAVs, referred to as mobile access points (MAPs) $\mathcal{L} = \{1, \ldots, L \}$, that are each placed at a height of $H_i \in \mathbb{R^+}, i = \{1, \ldots, L\}$. The Cartesian coordinates of the MSDs at time $t$ are denoted by $\boldsymbol{y}(t) = [y_1(t),y_2(t),\ldots,y_M(t)]^T$, where $y_i(t) \in \mathbb{R}^3, \forall i \in \mathcal{M}, t \geq 0$. Similarly, the Cartesian coordinates of the MAPs at time $t$ are denoted by $\boldsymbol{q}(t)=[q_1(t),q_2(t),...,q_L(t)]^T$, where $q_i(t) \in \mathbb{R}^3, \forall i \in \mathcal{L},t \geq 0$. The velocity of the MAPs at time $t$ are denoted by $\boldsymbol{p}(t)=[p_1(t),p_2(t),...,p_L(t)]^T$, where $p_i(t) \in \mathbb{R}^3, \forall i \in \mathcal{L},t \geq 0$. We assume that the MSDs are partitioned into $K \in \mathbb{Z}^+$ geographically separated sets $\mathcal{S} = \{S_1,S_2,...,S_K\}$ and the centroid of each set or cluster $S_i$ is denoted by $C_i$.

The MAPs have a maximum communication range of $r \in \mathbb{R}^+$ such that any two MAPs can communicate only if the Euclidean distance between them is less than $r$. The communication neighbours of each MAP is represented by the set $N_i = \{j \in \mathcal{L},j \neq i : \|q_i - q_j\| \leq r \},\forall i \in \mathcal{L}$. The quality or strength of the communication links between the MAPs is modeled using a distance dependent decaying function $\alpha_{z_1 ,z_0} (z) \in [0,1]$ with finite cut-offs, expressed as follows~\cite{flocking_oa}:
\vspace{-0.00in}
\begin{multline}
    \alpha_{\{z_1,z_0\}}(z)=\\
    \left\{  
        \begin{array}{ll}  
        1,& \text{if}\ 0\leq z<z_1,\\  
        \frac{1}{2}\left(1+\cos(\pi\frac{z-z_1}{z_0-z_1})\right),& \text{if} \ z_1\leq z<z_0,\\  
        0,& \text{if} \ z\geq z_0,   
        \end{array}  
    \right.
\end{multline}
where $z_0$ and $z_1$ are constant cut-offs.

As the MSD to MAP path-loss \cite{suburban_path_loss} is distance dependent, we assume an MSD $i$ always establishes connection with the MAP offering best quality of service:
\begin{equation}
    j = \underset{k\in \mathcal{L}:\|y_i-q_k\| \leq r}{\arg\max}\rho\|y_i-q_k\|^{-\eta},
    \label{eq:match}
\end{equation}
where $\rho \in \mathbb{R^+}$ is the transmission power and $\eta \in \mathbb{R^+}$ is the pass-loss exponent. The number of MSDs connected to each MAP is denoted by $N_u^k,\ k \in \mathcal{L}$.

\subsection{MAP Dynamics}

We leverage the widely accepted kinematic model in robotics and control literature to describe the dynamics of the MAPs as follows:
\begin{equation}
    \begin{aligned}
        \dot{q_i}=p_i, \\ 
        \dot{p_i}=u_i,
    \end{aligned}
\end{equation}
where $q_i,p_i,u_i \in \mathbb{R}^3$ and $i \in \mathcal{L}$. 
The control input can be designed to consist of the following three terms:
\begin{equation}\label{eq:4}
    u_i=f_i+g_i+h_i,
\end{equation}
where $f_i$ is an inter-MAP attractive/repulsive term, $g_i$ is a velocity consensus term, $h_i$ is a term defining the individual target of each MAP.

\subsection{Coverage and Connectivity}
The network coverage can be determined by the proportion of MSDs in the network that are successfully connected to an MAP. It is referred to as the coverage ratio of the network, $R_c$ and can be determined as follows:
\begin{equation}
    R_c = \frac{1}{M} \sum_{i=1}^{L} N_u^i,
\end{equation}
It reflects the formation status of MAPs and is used as a criteria in the mode switching.

The connectivity of the MAP network can be measured in terms of its Fiedler value (i.e., the second-smallest eigenvalue of the Laplacian matrix $\boldsymbol{L}$)~\cite{coverage_connectivity_tradeoff}. The adjacency matrix $\boldsymbol{A}=[a_{i,j}]$ is defined as:
\begin{equation}
    a_{i,j}=
    \left\{  
        \begin{array}{ll}  
        1,& \text{if} \ \|q_i-q_j\| \leq r,\\
        0,& \text{otherwise},   
        \end{array}  
    \right.
\end{equation}
where $i,j \in \mathcal{L}$. The Laplacian matrix of the MAP network is defined as $\boldsymbol{L}=\boldsymbol{D}-\boldsymbol{A}$ with $\boldsymbol{D}$ as the degree matrix of $\boldsymbol{A}$.
The Fiedler value is non-zero if each MAP in the network is reachable from any of the other MAPs, and hence provides a useful measure of global network connectivity. Furthermore, the higher the Fideler value, the more robust and resilient the network will be from a connectivity standpoint.

\section{Methodology} \label{Sec:Methodology}
This section presents the methodology used to develop the coverage formation and the connectivity establishment for MAPs. We create a resilient and autonomous configuration of MAP network through a multi-modal coverage control algorithm.

\subsection{Multi-modal Coverage Control}
In order to create tailored formations of MAPs to cover the MSDs in the network, we need to create several different modes of operation for the MAPs. Assume the centroids $\mathcal{C}=\{C_1,C_2,...,C_k\}$ of all MSD clusters $\mathcal{S} = \{S_1,S_2,...,S_k\}$ are known and the MAPs keep a minimum-weight spanning tree (MST) of MSD clusters that are already covered $\mathcal{T}_c=\{S_c^1,S_c^2,...,S_c^k\}$. At any given time, each MAP can be in one of the following operation modes:
\begin{itemize}
    \item[$M_0$:] When in Dynamic mode, the MAP sets its goal to the nearest cluster center $C_i$ that is not covered and travels to the cluster.
    \item[$M_1$:] When in Connectivity mode, the MAP determines the next cluster to connect and establish connectivity based on distributed MST algorithmn \cite{distributed_dijkstra}.
    \item[$M_2$:] When in Static mode, the MAP stays and serves MSDs in its goal cluster.
\end{itemize}
The initial modes of all MAPs are set to $M_i(0)=M_0$. In each iteration, each MSDs match with the nearest MAP using \eqref{eq:match}, and the MAPs share information such as relative position, velocity, number of connected MSDs, etc. with their neighbors. Then the network-wide coverage ratio is used for the switching modes and computing dynamics. The control algorithm is shown in Algorithm~\ref{alg:map_control}.
\begin{algorithm}
	\caption{MAP Control}
	\label{alg:map_control}
	\begin{algorithmic}[1]
	\Require Initialize position, velocity and mode for each MAP $q_i(0)$, $p_i(0)$, $M_i(0)\gets$$M_0$.
	\While {not converged}
	\State Determine the number of connected MSDs for each MAPs ($N^i_u(k)$).
	\State Each MAPs share the position, velocity, number of connected MSDs, list of achieved goals with its neighbors.
	\State Determine network-wide coverage ratio.
	\State Each MAP updates mode $M(k)$ using Algorithm \ref{alg:mode_switch}.
	\State Compute control input $u_i(k)$ for each MAP using~\eqref{eq:9}.
	\State Update the position and velocity of each MAPs using the discretized MAP dynamics.
    \EndWhile
	\end{algorithmic} 
\end{algorithm}

\vspace{-0.0 in}

\subsection{Connectivity Potential}
In order to form connectivity between clusters, we design a connectivity potential based on positions of cluster centroids. Suppose the MAP $i$ connects clusters $S_1$ and $S_2$ with centroids $C_1$ and $C_2$ respectively. In practice, the cluster centers can be determined by an aerial survey of the ground population and users. Several techniques such as simultaneous localization and mapping (SLAM) and other imaging technologies are now available that can automate the process to provide information about the key centers of network users.
The connectivity potential function is defined as:\vspace{-0.00in}
\begin{equation}
    E_c^i(q)= k(\|q_1^r-q_i\|_\sigma+\|q_2^r-q_i\|_\sigma-\|q_1^r-q_2^r\|_\sigma).
\end{equation}
where the $\sigma$-norm, $\|z\|_\sigma$ is defined as\vspace{-0.00in}
\begin{equation}
    \|z\|_\sigma=\frac{1}{\epsilon}(\sqrt{1+\epsilon\|z\|^2}-1),
\end{equation}\vspace{-0.0in}
with $\epsilon>0$ a positive constant and the gradient can be expressed as
\begin{equation}
    \nabla \|z\|_\sigma=\frac{z}{\sqrt{1+\epsilon \|z\|^2}}=\frac{z}{1+\epsilon\|z\|_\sigma}.
\end{equation}
The advantage of $\|z\|_\sigma$ is that it is differentiable everywhere while traditional norm $\|z\|$ is not differentiable at $z=0$. This creates a smooth potential function.

\begin{figure}[t]
    \centering
    \includegraphics[trim={0 0 0 0.6in},clip,width=3in]{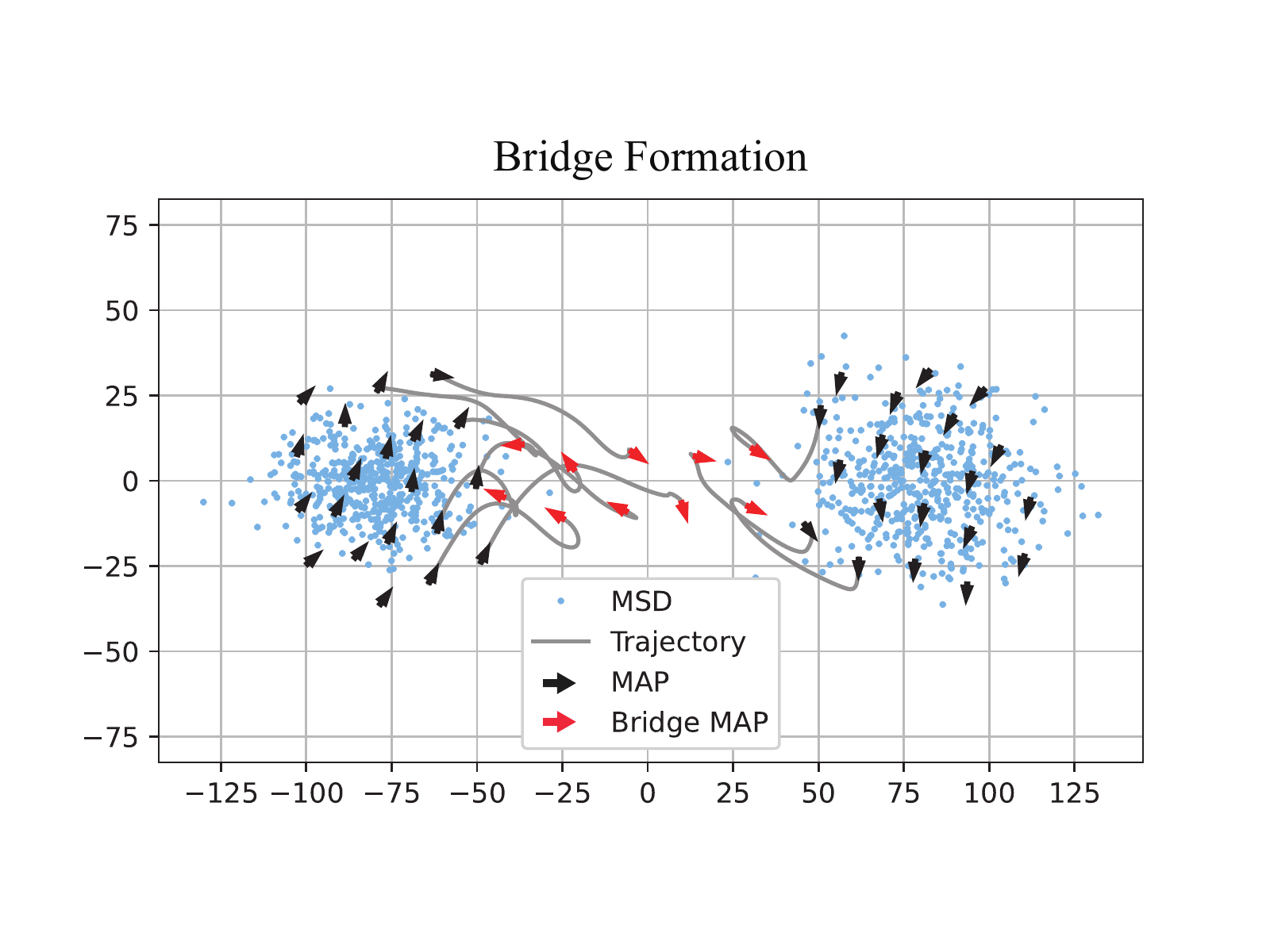} \vspace{-0.2in}
    \caption{An example of trajectories of MAPs connecting two clusters.\vspace{-0.1in}}
    \label{fig:traj}
\end{figure}

The effectiveness of the connectivity potential is shown in Fig. \ref{fig:traj}. Free MAPs can form a bridge between two clusters and provide connectivity.

\begin{figure*}[t]
     \centering
     \begin{subfigure}{0.24\linewidth}
         \centering
         \includegraphics[width=\linewidth]{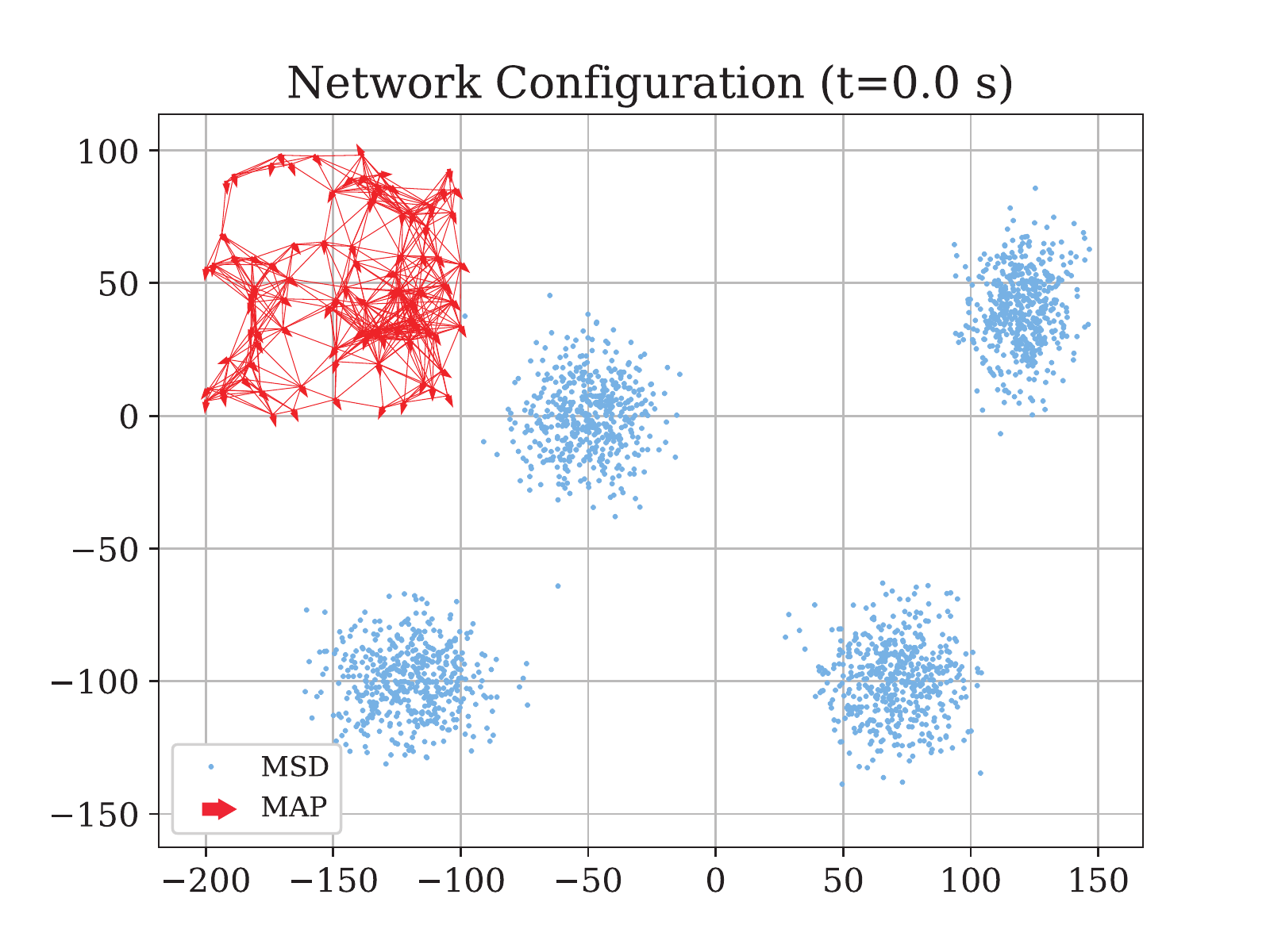}
         \caption{}
         \label{fig:a}
     \end{subfigure}
     \begin{subfigure}{0.24\linewidth}
         \centering
         \includegraphics[width=\linewidth]{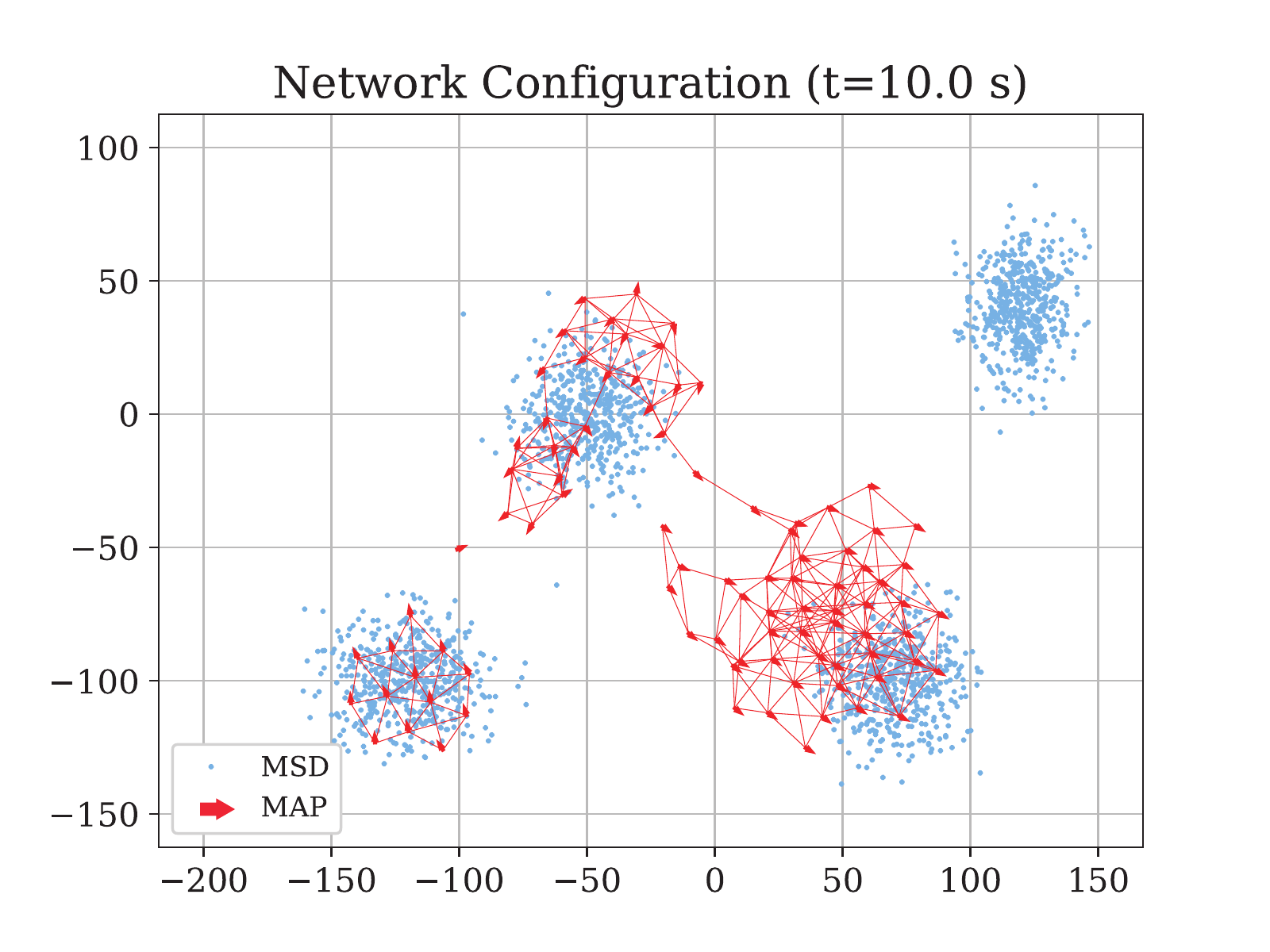}
         \caption{}
         \label{fig:b}
     \end{subfigure}
     \begin{subfigure}{0.24\linewidth}
         \centering
         \includegraphics[width=\linewidth]{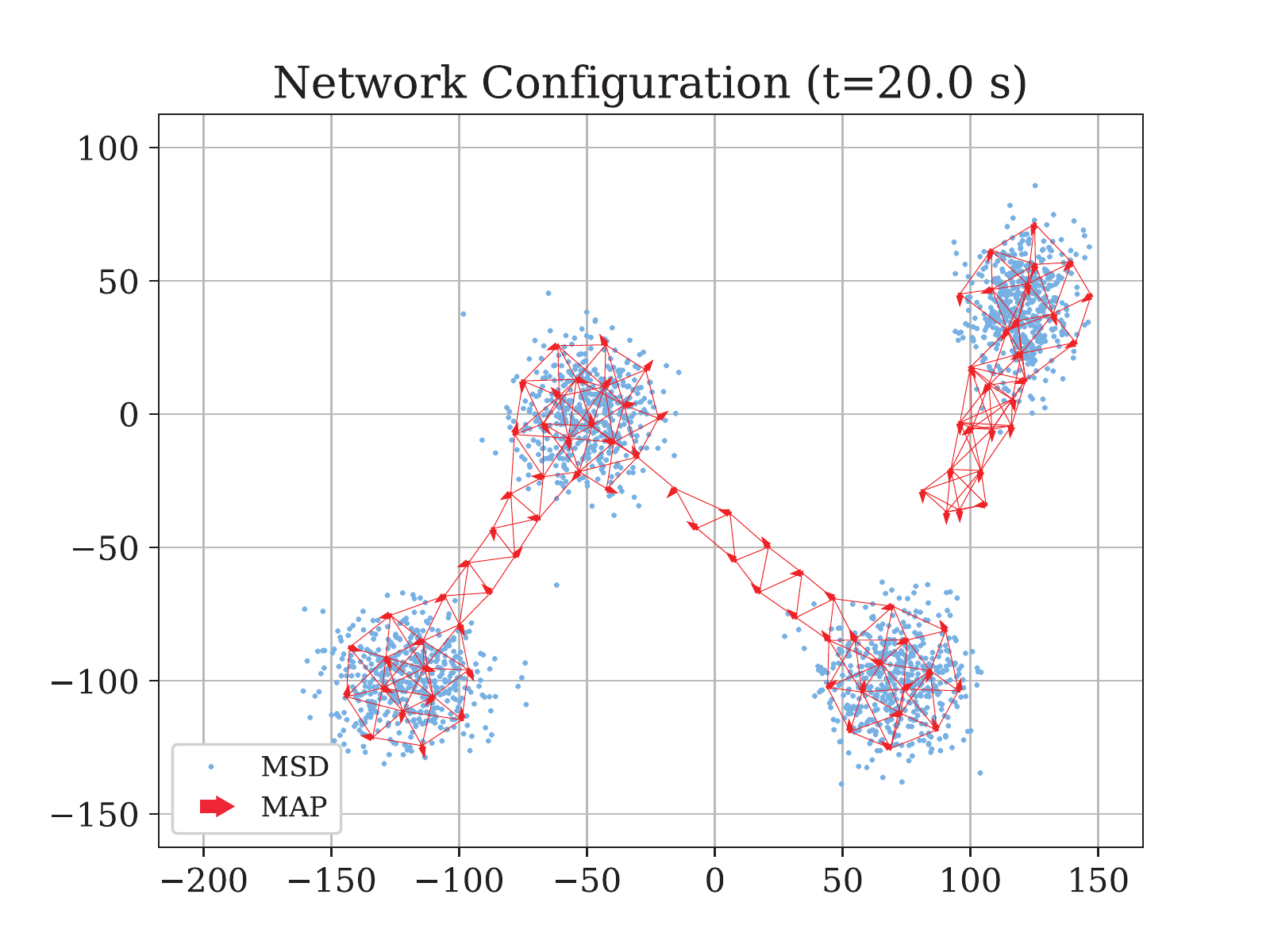}
         \caption{}
         \label{fig:c}
     \end{subfigure}
     \begin{subfigure}{0.24\linewidth}
         \centering
         \includegraphics[width=\linewidth]{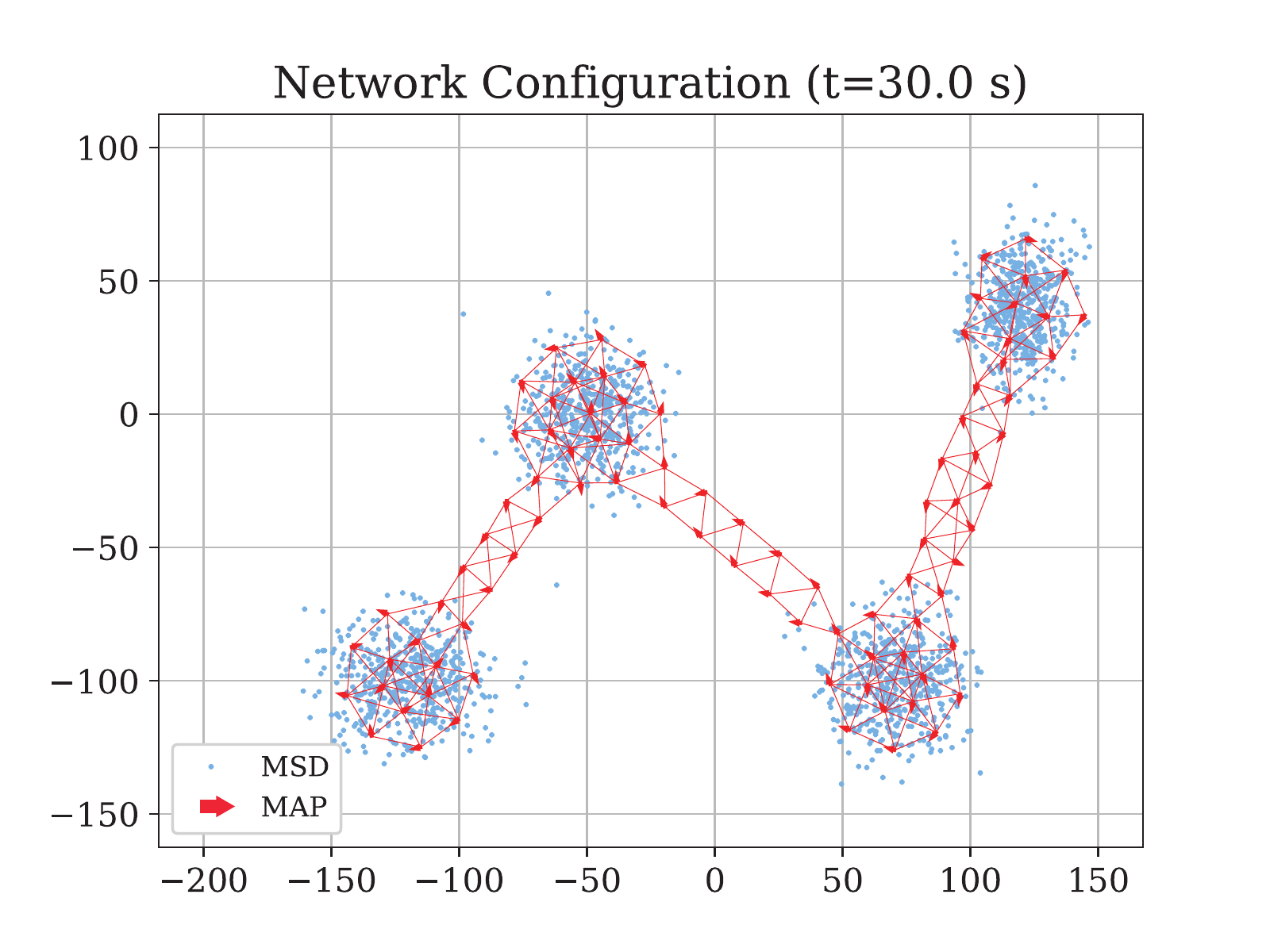}
         \caption{}
         \label{fig:d}
     \end{subfigure}
    \caption{Experiment results for network formation and connectivity. Figure (a) shows the initial positions of MAPs and MSDs. Figures (b), (c) and (d) show the process of MAPs covering all four MSD clusters and creating connectivity between them.\vspace{-0.1in}}
    \label{fig:network configuration}
\end{figure*}

\subsection{MAP Dynamics}
The control input $u_i$ can be designed to consist of three components as follows \cite{resilient_connectivity_tccn}:
\begin{equation}\label{eq:9}
    u_i=f_i(q,A,N_u)+g_i(p,A)+h_i(q,p),
\end{equation}
where $f_i(q,A,N_u)$ defines the gradient based attractive and repulsive term, $g_i(p,A)$ defines the velocity consensus term and $h_i(q,p)$ defines the goal approach term.

\begin{enumerate}
    \item Attractive and repulsive functions:
        \begin{multline}
            f_i(q,A,N_u)=\sum_{j\in N_i}\Bigg[\Phi (\|q_j-q_i\|_\sigma)+\\ a\left(1-\alpha_{\{0,1\}}\left(\frac{\|(N_u^j-N^{max})^+\|_\sigma}{\|N^{max}\|_\sigma}\right)\right) \Bigg]\textbf{v}_{i,j},
        \end{multline}
        where $\textbf{v}_{i,j}=\nabla \|q_j-q_i\|_\sigma$ is the vector from $q_i$ to $q_j$. The function $\Phi(z)$ is expressed as:
        \begin{equation}
            \Phi(z)=\alpha_{\{\gamma,1\}}\left(\frac{z}{\|r\|_\sigma}\right)\phi(z-\|d\|_\sigma),
        \end{equation}
        where $\phi (z)=\frac{1}{2}[(a+b)\frac{(z+c)}{\sqrt{1+(z+c)^2}}+(a-b)]$ and $c=|a-b|/\sqrt{4ab}$ to ensure that $\phi (0)=0$. Here $r$ is the maximum communication range and $d$ is the minimum distance between MAPs.
    \item Velocity consensus function:
        \vspace{-0.0in}
        \begin{multline}
            g_i(p,A)=\\ \sum_{j\in N_i \backslash i }\left(1-\alpha_{\{0,1\}}\left(\frac{\|(N^{max}-N_u^i)^+\|_\sigma}{\|N^{max}\|_\sigma}\right)\right)\\
            a_{ij}(p_j-p_i).
        \end{multline}
        The velocity concensus function works as a damping force that leads to a match bewteen neighboring MAPs. This can reduce potential collisions and disconnections between MAPs.
    \item Goal functions:\\
    We define a goal function besed on the MAP mode $M_i(t)$. It shows a tendency to approach a dynamic/static group objective.
        \begin{enumerate}
            \item Goal function (for Mode $M_0$ and $M_2$):
                \begin{equation}\label{eq:12}\hspace{-0.01in}
                    h_i(q,p)=c_1(q_i^r-q_i)+c_2(p_i^r-p_i).
                \end{equation}
            \item Goal function (for Mode $M_1$):
                \begin{multline}\label{eq:13}\hspace{-0.14in}
                    h_i(q,p)=\nabla E_c(q)+\frac{1}{2}c_2(p_1^r-p_i)+\frac{1}{2}c_2(p_2^r-p_i)\\
                    =\frac{k(q_1^r-q_i)}{1+\epsilon\|q_1^r-q_i\|_\sigma}+ \frac{k(q_2^r-q_i)}{1+\epsilon\|q_2^r-q_i\|_\sigma}\\ +\frac{1}{2}c_2(p_1^r-p_i)+\frac{1}{2}c_2(p_2^r-p_i).
                \end{multline}
        \end{enumerate}
\end{enumerate}

\begin{algorithm}
	\caption{Mode Switch}
	\label{alg:mode_switch}
	\begin{algorithmic}[1]
	\Require Current mode $M_i(k)$, Number of served MSDs $N^i_u(k)$, coverage ratio $r^i_g(k)$ for current goal $g^i_c(k)$, coverage thresholds $r_0$, serving capacity thresholds $n_0$ and $n_1$.
	\If {$r^i_g(k)>$ $r_0$}
	    \State Update MST of achieved goals $\mathcal{T}_c \gets \mathcal{T}_c + g^i_c$.
	    \If {$M_i(k)=M_0$}
    	    \If {$0 < N^i_u(k)< $ $n_0$}
    	        \State Determine the nearest cluster center $g_u^i$ that $g_u^i\notin \mathcal{T}_c$.
    	        \State Update current goal $g_c^i \gets g_u^i$.
    	    \ElsIf {$n_0 \leq N^i_u(k)< $ $n_1$}
    	        \State Update current mode $M_i(k)\gets M_1$.
    	        \State Determine the clusters to connect using distributed MST algorithm \cite{distributed_dijkstra} and build connectivity using the controller in Equation \eqref{eq:9}.
    	    \Else 
    	        \State Update current mode $M_i(k) \gets M_2$
    	        \State Serve MSDs in current goal cluster using the controller in \eqref{eq:9} and \eqref{eq:12}
    	    \EndIf
    	\Else
    	    \State $M_i(k) \gets M_i(k)$
	    \EndIf
    \Else
        \State $M_i(k) \gets M_i(k)$
	\EndIf
	\end{algorithmic} 
\end{algorithm}

\subsection{Backhaul Connectivity Algorithm}
 We create a backhaul connectivity algorithm based on MAP modes $M_i(t)$. Suppose the centroids $C_i$ of all MSD clusters are known. All MAPs are initialized with random position $q_i(0)$, velocity $p_i(0)$ and mode $M_i(0)=M_0$. Then each MAP switches mode from $M_i(k)$ to $M_i(k+1)$ based on current mode $M_i(k)$, number of MSDs it serves $N_u^i(k)$ and the coverage ratio for current goal $r_g^i(k)$. When the MAPs switch to $M_1$, they will establish connectivity between clusters using the connectivity functions defined in Equation \ref{eq:9}. The detailed mode switch algorithms is defined in Algorithm~\ref{alg:mode_switch}.

\section{Simulation Results} \label{Sec:Results}

In this section, we demonstrate the effectiveness of our proposed solution with simulations in Python platform. MAPs are released from a uniformly distributed area centered at (-150,50) wih fixed height $h_i=20$ m. The initial velocity of MAPs are randomly selected from $[-1,1]^2$. The MSDs are divided into four clusters using 2D Gaussian distribution and each cluster has 500 MSDs for all simulations. The following parameters persist throughout the experiments: minimum separation between MAPs $d=20$ m, communication range of MAPs $r=1.2d,\ \epsilon=0.1 $ for $\|z\|_\sigma,$ transmit power $\rho = 1\ \text{W}$, path-loss exponent between MAPs and MSDs $\eta = 3.5$, $\ a=b=5$ for $\phi(z),\ N_{max}=80,\ c_1=0.3,\ c_2=0.6,\ k=10 $ for goal functions$,\ r_0=0.95,\ n_0=3,\ n_1=10$ for mode switching, Simulation time step $\Delta t=0.1\ s.$

In Fig. \ref{fig:network configuration}, we show an example of the experiment results for our proposed method using 90 MAPs. Fig. \ref{fig:a} shows the initialization of the MAPs and MSD. The MAPs traverse the four clusters, build connectivity between clusters and serve MSDs in their individual goal clusters. Finally at $t=30.0$ $s$ as shown in Fig. \ref{fig:d}, the network converge and develop a connected network covering all four clusters.

\begin{figure}[t]
    \centering
    \includegraphics[trim={0 0 0 0.5in},clip,width=2.9in]{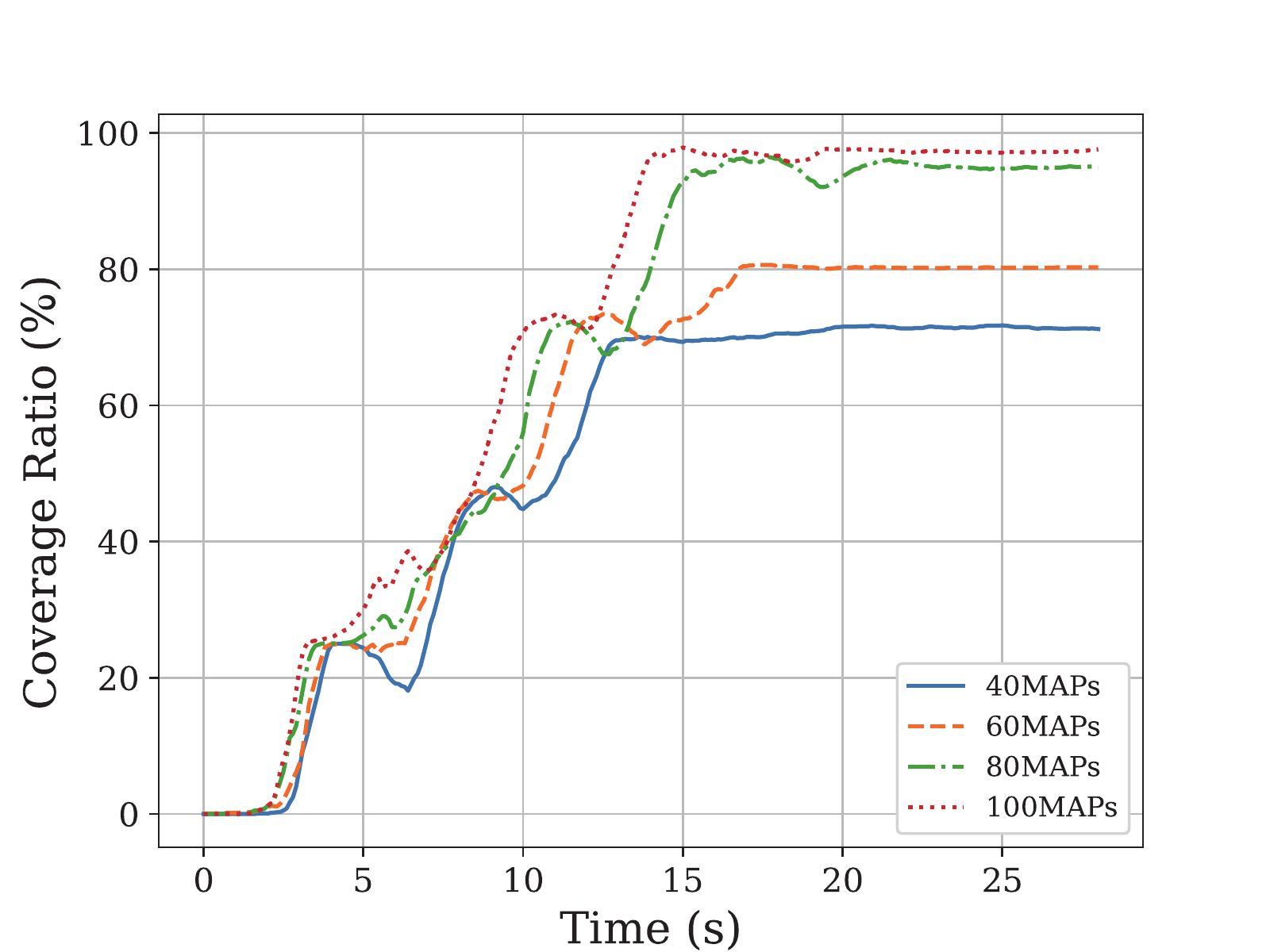}
    \caption{Coverage ratios with time in experiments using different numbers of MAPs. The highest coverage ratio using 100 MAPs reaches $97.3\%$ after convergence.\vspace{-0.1in}}
    \label{fig:f}
\end{figure}

\begin{figure}[t]
    \centering
    \includegraphics[trim={0 0 0 0.5in},clip,width=3.0in]{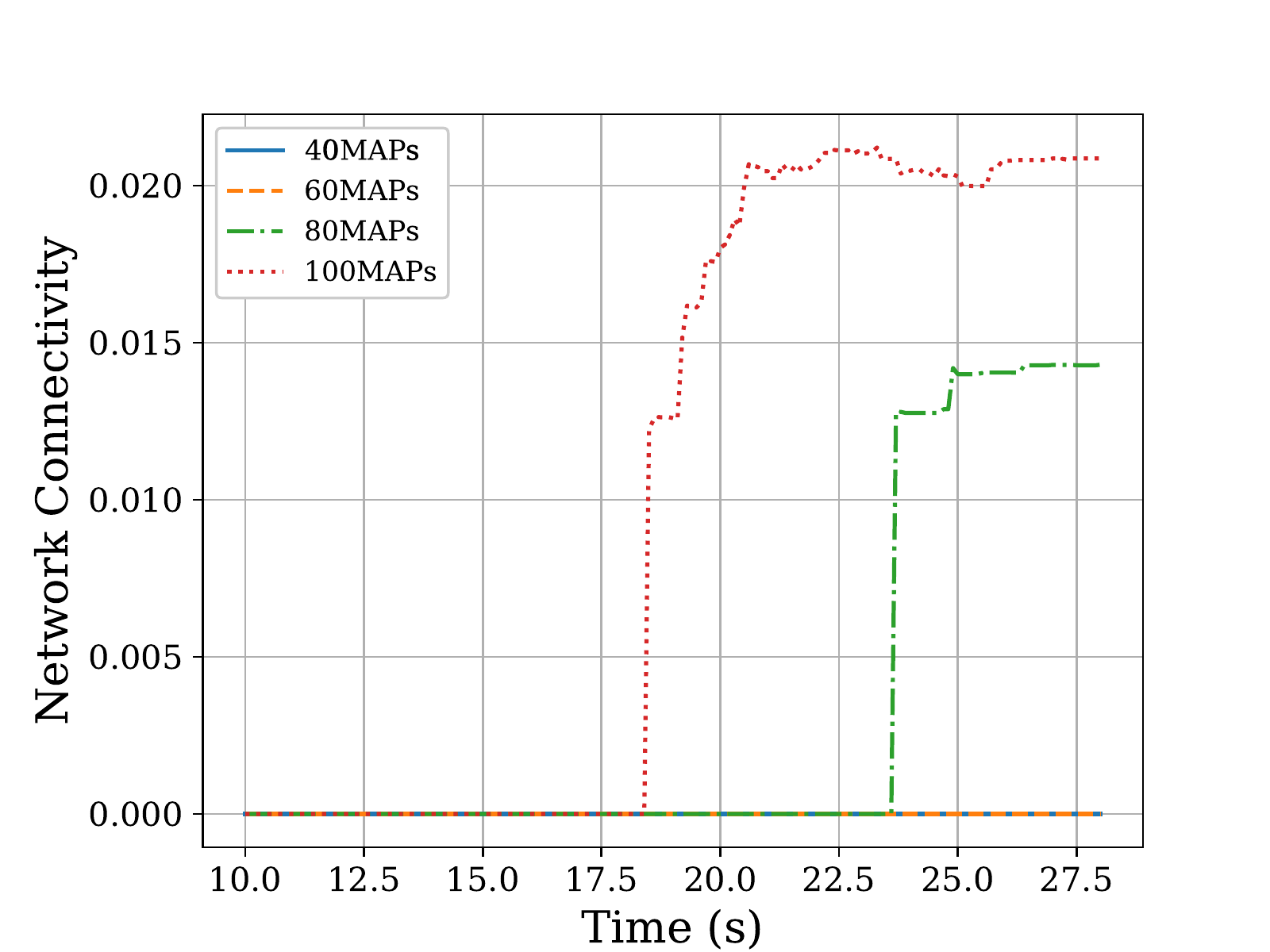}
    \caption{Fiedler values with time in experiments using different numbers of MAPs. The highest Fiedler value using 100 MAPs reaches 0.021 after convergence.\vspace{-0.0in}}
    \label{fig:g}
\end{figure}

In Fig. \ref{fig:f} and Fig. \ref{fig:g}, we compared the coverage ratios and Fiedler values using different number of MAPs. In situation when only 40 MAPs are deployed, the MAPs are unable to cover the fourth cluster due to the limitation of MAP quantity. The coverage ratio only reaches $70\%$ after convergence. In situations when 40 or 60 MAPs are deployed, the Fiedler values remain zero after convergence because the network has insufficient MAPs to build connectivity among all clusters. In comparison, when 80 or 100 MAPs are deployed, the MAPs can form a desired connected network which cover all clusters with coverage ratios over $95\%$.

\begin{figure}[t]
    \centering
    \includegraphics[trim={0 0 0 0.5in},clip,width=3.0in]{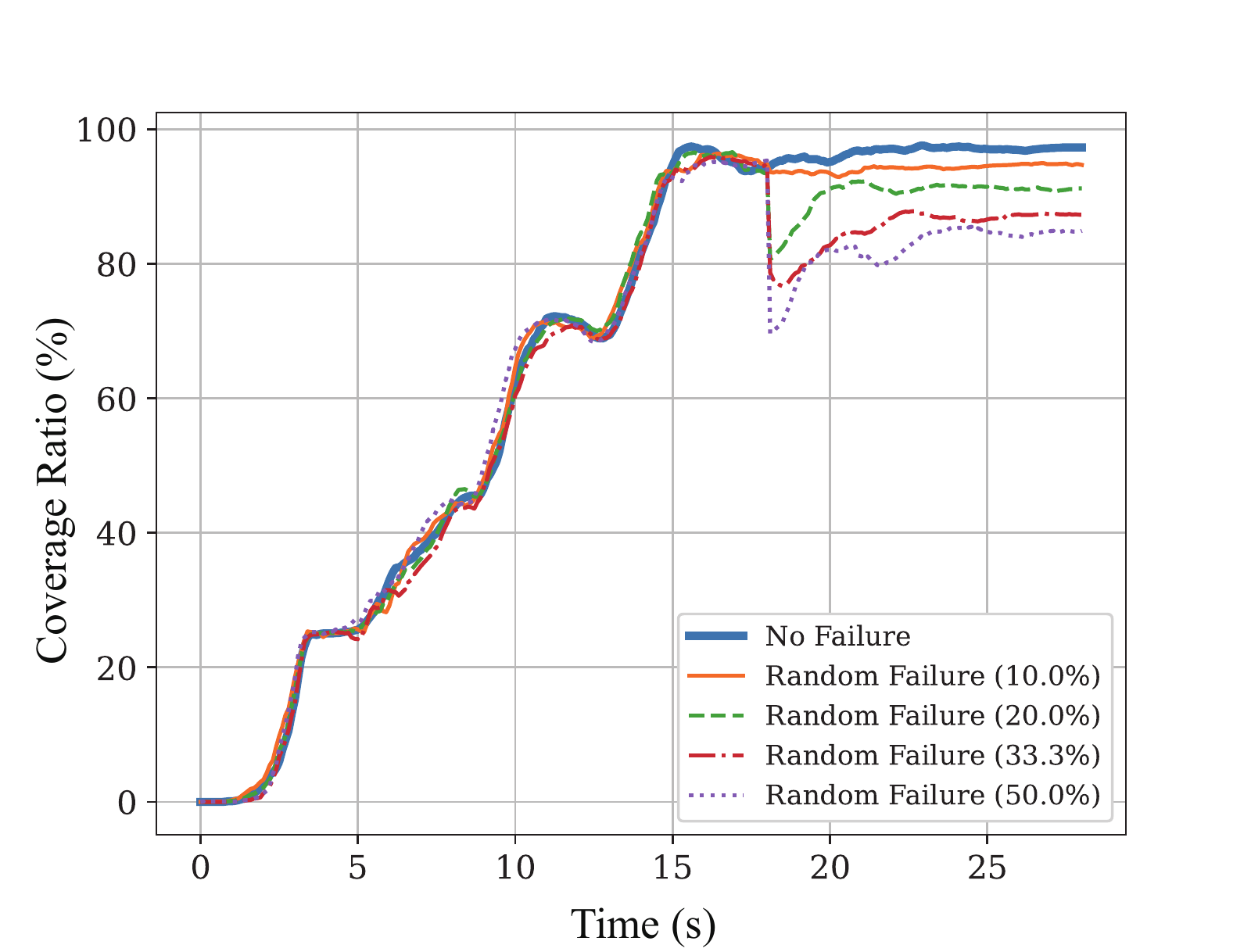}
    \caption{Coverage ratios with time when random failure happens at $t=18 s$.\vspace{-0.1in}}
    \label{fig:h}
\end{figure}

In Fig. \ref{fig:h}, we analyze the resilience of the network when random failures of MAPs occur. 80 MAPs are used to build the network and random MAPs stop working at $t=18s$ with failure ratios from $10.0\%$ to $50.0\%$. With no failure happening, the coverage ratio keeps increasing until it reaches a highest value of $97\%$. When failure occurs at $t=18s$, the MAPs react quickly and restore coverage ratios. However, when the failure ratio is $50\%$, although the MAPs are still able to maintain a coverage ratio of $85\%$, the connectivity among clusters is lost.

\begin{figure}[t]
    \centering
    \includegraphics[trim={0 0 0 0in},clip,width=2.8in]{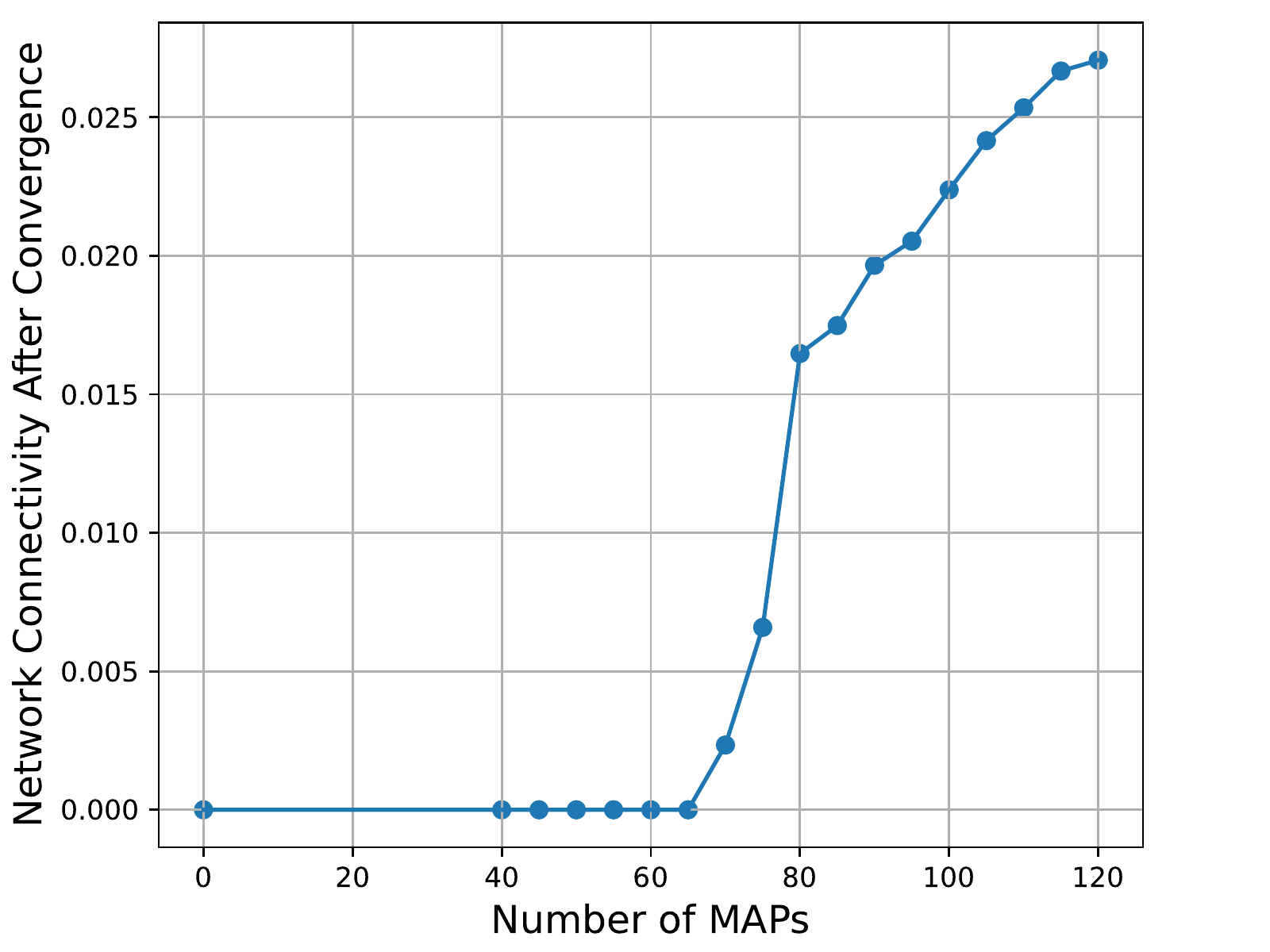}
    \caption{Fiedler values after network convergence using different number of MAPs.\vspace{-0.0in}}
    \label{fig:i}
\end{figure}

In Fig. \ref{fig:i}, we compare connectivity of the network (measured from Fiedler values after convergence) using different number of MAPs. When the number of MAPs is less than 65, the MAPs are unable to provide connectivity among all four clusters. So the Fiedler value in this case is zero. When the number of MAPs is between $(65,80)$, the Fiedler value increases drastically as the number of MAPs increases. When the number of MAPs is in the interval of $(80,120)$, the Fiedler value keeps increasing, however, the rate of increase reduces gradually.

\section{Conclusions} \label{Sec:Conclusion}
In this paper, we presented a potential approach to construct a mobile aerial network that provides both local coverage and back-haul connectivity. The proposed method was inspired from flocking and distributed swarming behaviours in nature. The experimental results showed that the solution has been successful in maintaining a stable and high coverage ratio after the network converges. The designed connectivity algorithm was able to form communication bridges between spatially dispersed clusters. Further more, the network had resilience to random failures and self-recovery ability. Once failures occurred, the working UAVs can autonomously reconfigure the network and restore coverage and connectivity. In cases of extremely high failure ratios, the network may lose connectivity but can still provide local coverage around the cluster centers. Future work will focus on adding quality-of-service based coverage and connectivity to provide differentiated services in aerial networks.

\ifCLASSOPTIONcaptionsoff
  \newpage
\fi

\bibliographystyle{IEEEtran}
\bibliography{IEEEabrv,Bibliography}

\end{document}